\documentclass{birkjour}
 \theoremstyle{definition}
 
 \theoremstyle{remark}

 \numberwithin{equation}{section}
\newcommand{\Pin}{\mathop{\mathrm{Pin}}}
\newcommand{\Spin}{\mathop{\mathrm{Spin}}}

\usepackage{lineno}
\usepackage{tikz}
\usepackage{pgfplots}
\usepgflibrary{shapes.geometric} % LTEX and plain TEX and pure pgf 
\usepgflibrary[shapes.geometric] % ConTEXt and pure pgf 
\usetikzlibrary{shapes.geometric} % LATEX and plain TEX when using Tik Z 
\usetikzlibrary[shapes.geometric] % ConTEXt when using Tik Z 

\begin{document}
	\bibliographystyle{plain} 
%-------------------------------------------------------------------------
% editorial commands: to be inserted by the editorial office
%
%\firstpage{1} \volume{228} \Copyrightyear{2004} \DOI{003-0001}
%\seriesextraline{This is the Concrete Title of this Book\br H.E. R and S.T.C. W, Eds.}
%\firstpage{1}

\title[From symmetries of viruses to $E_8$ \& ADE correspondences]
 {Root systems \& Clifford algebras: from symmetries of viruses to $E_8$ \& ADE correspondences}

%----------Author 1
\author[P-P Dechant]{Pierre-Philippe Dechant}

\address{%
Departments of Mathematics and Biology\\
York Cross-disciplinary Centre for Systems Analysis\\
University of York, Heslington YO10 5GE, United Kingdom}

\email{ppd22@cantab.net}



\date{August 11, 2017}
\dedicatory{%Proceedings of the Nankai Symposium on Geometry, Physics and Number Theory 2017 \\ 
Dedicated to the memory of the late Isadore Singer}
%%% ----------------------------------------------------------------------

\begin{abstract}
In this paper we discuss reflection groups and root systems, in particular non-crystallographic ones,
and a Clifford algebra framework for both these concepts. A review of historical as well as more recent work on viral capsid symmetries motivates 
the focus on the icosahedral root system $H_3$. We discuss a notion of affine extension for non-crystallographic 
groups with applications to fullerenes and viruses. The icosahedrally ordered component of the nucleic acid
within the virus capsid and the interaction between the two have shed light on the viral assembly process with interesting applications to 
antiviral therapies, drug delivery and nanotechnology. The Clifford algebra framework is very natural, as it uses precisely the structure that is already implicit in this root system and reflection group context, i.e. a vector space with an inner product. In addition, it affords a uniquely simple reflection formula, a double cover of group transformations, and more insight into the geometry, e.g. the geometry of the Coxeter plane. This approach made possible a range of root system induction proofs, such as the constructions of $E_8$ from $H_3$ and  the exceptional 4D root systems from 3D root systems. This makes explicit various connections between what Arnold called Trinities (sets of three exceptional cases). In fact this generalises further since the induction construction contains additional cases, namely two infinite families of cases. It therefore actually yields $ADE$ correspondences between three sets of different mathematical concepts  that are usually thought of separately as polytopes, subgroups of $SU(2)$ and $ADE$ Lie algebras. Here we connect them explicitly and in a unified way by thinking of them as three different sets of root systems. 

\end{abstract}
%%% ----------------------------------------------------------------------
\maketitle
%%% ----------------------------------------------------------------------
%\tableofcontents

\section{Introduction}\label{intro}
Polyhedral symmetries are ubiquitous, and occur frequently when larger structures are built 
from identical subunits. Examples are viruses, carboxysomes, radiolaria and diatoms in biology,
fullerenes and salt crystals in chemistry, and even geodesic domes in architecture. 
These polyhedral symmetries are the symmetries of the Platonic solids, the regular convex polytopes.
They are generated by reflections, even if ultimately -- for instance in biology -- the relevant 
symmetries are only the subset of the rotations. 

We are therefore firstly interested in symmetry groups generated by reflections; in particular, root systems and 
their simple roots are convenient ways to generate these reflection groups. Secondly, many of the above examples such as viruses
have icosahedral symmetry, which is an example of a non-crystallographic reflection group. These groups
are conventionally less studied when one comes from a Lie theory background, as they do not arise in this context by construction. Thirdly, 
Clifford algebras are a convenient framework for discussing reflection groups for a variety of reasons that we will discuss in the second half of the paper. Applying this Clifford framework to root systems and reflection groups  led to a range of deeper insights into supposedly familiar and well-understood topics as well as new results. The first part of this article is thus a review of applications of group theory to viruses and fullerenes, both historical and contemporary. The second half will discuss the Clifford algebraic framework and  new insights this has led to. 

\section{Viruses, root systems and affine extensions}\label{parti}
This section will discuss the reasons for viral symmetries and applications of group theory in the form of tilings, affine extensions and representation theory. Other application areas include architecture, virus-like nanoparticles, viral assembly and fullerene structure. 

\subsection{Viruses}\label{viruses}

Viruses are biological entities that transport genetic information (in the form of an RNA or DNA genome) between cells, hosts and even species. 
Since nucleic acid is fragile, it is protected by a protein shell called a capsid during this transport.
Because proteins are made from genes, viruses have the difficulty of having to encode the protein for this capsid within the genome contained within.
If the capsid were made from a single protein, the genome needed to code for this protein would be so large as not 
to fit inside. 
More generally, this encoding of protein information in a genome is only economical if biological systems use multiple copies of the protein gene product (at least on average). 
A good way of using multiple copies of the same protein building block is by using symmetry, and simple viruses as the smallest biological systems do exactly that.
For instance, Hepatitis B virus (HBV) and the virus MS2 each have 4 genes, of which only one is structural, corresponding to a single protein building block for the capsid.

The genome consists of a four-letter alphabet, can be either DNA or RNA, and single- or double-stranded. The single-stranded
versions are floppy and more fragile, and form complex secondary structures, e.g. by folding back on themselves and Watson-Crick basepairing with themselves. The double-stranded versions have two complementary strands Watson-Crick basepaired, and are rather more stable, providing a very stable genetic information carrier for large information volumes -- e.g. humans have dsDNA.
Three-letter words (codons) in this four-letter alphabet (i.e. $64$ possibilities) each code for one of $20$ amino acids, i.e. there is some redundancy in this code.
The different codons in the gene are read and the corresponding amino acids are strung together into a polypeptide chain like beads on a string. 
However, like ssNA, this chain begins to fold and form secondary and tertiary structures, forming a specific geometric shape in 3D with certain interaction forces around the edges. Again there is therefore a large degree of degeneracy, since different amino acid chains might result in a very similar folded state, and often small amino acid substitutions do not have a large effect on the overall fold. The level of evolutionary selection probably largely applies on this level of tertiary structure functionality as a geometric/biophysical building block, rather than genetic or amino acid sequence as such \cite{abrescia2010does}. We will revisit this point later. 

In the 1950s, early electron micrographs showed viruses to be either rod-shaped or spherical. Watson and Crick, however, realised that because all biological systems are ultimately made from discrete components, e.g. proteins, these structures can not be exactly rods or spheres, as their symmetry groups are not discrete \cite{Crick:1956}. 
Examples of discrete symmetry groups that are available are the polyhedral groups. Objects with polyhedral symmetry can appear roughly spherical, and in fact in the electron micrographs one might be able to make out a certain graininess reminiscent of the vertices of a polyhedron. Watson and Crick therefore conjectured on theoretical grounds that the spherical viruses should in fact be polyhedral, whilst the rod-like ones should be helical.
The largest of the available polyhedral groups is the icosahedral group, and thus icosahedral symmetry would appear to be evolutionarily and biologically the most advantageous. Subsequent experiments have of course confirmed this. 

The group of rotational symmetries of the icosahedron is of order $60$, and is called $I$, or $A_5$, in its guise as the alternating group in five elements. The full symmetry group including reflections is of order $120$ and is sometimes called $I_h$, though we will use the Coxeter notation $H_3$. This is a subgroup of $O(3)$ and doubly covers the rotational subgroup, which sits inside $SO(3)$. There is also a double cover in $\Spin(3)$, which will make an appearance later and which we call the binary icosahedral group $2I$. The icosahedral group has 5 conjugacy classes consisting of the identity, $15$ $2$-fold rotations around the edges, $20$ $3$-fold rotations around the faces, and two classes of $12$ $5$-fold rotations by $2\pi/5$ and $4\pi/5$, respectively. It thus has irreducible matrix representations of dimensions $1$, $3$, $3$, $4$ and $5$. The full icosahedral group $H_3=\mathbb{Z}_2\times A_5$ just has two copies, whilst the binary icosahedral group has four additional conjugacy classes and thus has an additional four irreducible representations of dimensions $2$, $2$, $4$ and $6$. The icosahedron consists of 20 equilateral triangles, such that if a virus can make 20 triangles it can successfully encapsidate its genome. The fundamental domain, however, is only one third of such a triangle (in fact an exactly equilateral triangle would be difficult to make from a single protein), such that viruses only need to be able to make 60 thirds-of-a-triangle (taken e.g. as a kite-shaped fundamental domain). One often denotes this protein in each fundamental domain by schematically putting a dot in each corner of the triangle. This gives rise to clusters of 5 proteins (pentamers) around the 12 vertices of the icosahedron. 

Whilst icosahedral symmetry is the largest available symmetry group, reusing the same protein only 60 times is very modest by biological standards. E.g. expression of the proteins actin, myosin or ubiquitin is vastly higher than that. 
Thus there is evolutionary pressure on viruses to somehow improve on these constraints of icosahedral symmetry that there can only be 60 symmetry equivalent copies of a protein. However, the difficulty lies in making a gene (recipe) for a protein of the right shape and edge interactions rather than in making many copies of this protein. If there is already such a recipe for a triangle then there is virtually no additional cost for making more than $20$ triangles, e.g. $80$. Caspar and Klug therefore had the idea that viruses could employ subtriangulations of the icosahedron: e.g. $4$ small triangles can make a large triangle, and then  20 of these large triangles  make a larger icosahedron \cite{Caspar:1962}. This is obviously more economical genetically; however, exact symmetry equivalence of these protein locations is lost. The proteins form several orbits of the icosahedral group, with triangles being `quasi-equivalent' tiles that are not symmetry equivalent but in `nearly' identical environments. The protein clusters that now form away from the 12 vertices of the icosahedron are all hexameric. This improves the number of copies of the capsid protein to $60T$, where $T$ is the number of orbits and also the number of small triangles per large triangle, called the triangulation number. Caspar and Klug found the  triangulations of the icosahedron, which are only possible for certain $T$-numbers given by $T=h^2+hk+k^2$, where $h$ and $k$ are non-negative integers. This leads to the predictions of $60T$ total protein subunits, with 12 pentameric vertices and $10(T-1)$ hexameric clusters on the faces. 

Whilst this was a good idea, based on the fact that the icosahedron is already made of triangles such that one could try to subtriangulate it, this obviously does not give all icosahedral tilings. For instance, we have seen that the icosahedron can be tiled with $60$ identical kites, so one could wonder if other tilings were allowed that are also quasiequivalent (identical tiles in near-identical environments) but not based on triangulations. For instance, in the 90s the Papova- and Papillomaviruses (e.g. Human Papillomavirus, HPV and Simian Virus SV40) were shown to defy the predictions of Caspar-Klug theory by displaying only $72$ pentamers, and no hexamers at all. This motivated a more general approach to icosahedral viral tilings in the work of Reidun Twarock \cite{Twarock:2005a, Twarock:2005c, Twarock:2004a, Twarock:2006a, ElSawy:2008}. For instance, of course the dodecahedron -- the dual of the icosahedron -- tiles the sphere with pentagons or pentagonal wedges, and the rhombic triacontahedron (essentially the superposition of an icosahedron and a dodecahedron) tiles the sphere with $30$ golden rhombuses (i.e. the ratio of the two lengths is the golden ratio $\tau=\frac{1}{2}(1+\sqrt{5})$). This idea of quasiequivalence as having a single tile shape is thus analogous to having a single tile shape in lattices, e.g. tilings of the plane. This viral tiling theory gives not only the location of proteins and protein clusters as orbits of  the icosahedral group, but also takes into account other biological information such as the number of proteins bound into dimers (rhombus) or trimers (triangles or kites) within the tiles as well as bonding behaviour between tiles (edge decorations). E.g. Pariacotovirus might be best described by a triangulation, whilst Poliovirus is better described in terms of kites, and MS2 as a rhombus tiling with $90$ rhombuses, even though all three consist of three icosahedral orbits and thus nominally correspond to $T=3$. 

However, the Polyoma- and Papillomaviruses were still elusive. The necessary generalisation in fact consists in also relaxing the condition of quasiequivalence: this means that there is not a single type of tile -- but more than one. This idea was also encountered in tilings of the plane (after centuries of Islamic art) by Roger Penrose, who considered tilings of the plane with long-range order but without translation invariance. There are many such examples of aperiodic tilings, e.g. with as few as two different types of tile, such as rhombuses and pentagones, or a thin and a thick rhombus. It turned out that combining the two tile shapes of kite and rhombus finally explained the structure of HPV and SV40 with its 72 pentameric clusters, albeit at the price of relinquishing quasiequivalence. 

This viral tiling theory was a fairly comprehensive treatment of viral surface organisation for simple viruses. In fact, HPV is alreay an example of a dsDNA virus; this class of viruses can have exceedingly complex genomes and thus have to rely on symmetry rather less. For instance, Poxviruses are less symmetric, if at all (probably bilateral); Herpesviruses are icosahedral but have many more structural proteins, around $100$ genes overall and a distinguished vertex; tailed bacteriophages (Caudovirales) combine aspects of helical and icosahedral symmetry in their tail and head structures, or even employ two different T-number organisations in the same head; the largest icosahedral virus, Mimivirus, has a distinguished vertex called a stargate via which it fuses, a T-number of around $1000$ and many genes, comfortably above the number of genes of small bacteria; and the even more recently discovered giant viruses are asymmetric,  even larger, have even larger genomes and thoroughly blur the previously clear-cut boundaries between viruses and bacteria \cite{abergel2015rapidly}.

However, since (small) viruses have evolved firstly to icosahedral symmetry, secondly to generalisations of it via tilings, it is an interesting question whether the evolutionary pressure is high enough to go further: does it go beyond merely the surface organisation via a compact symmetry to extend the symmetry group to something potentially non-compact, by relating different points inside the virus via a translation, given by a length scale in the problem such as the size of protein or nucleic acid? We will  consider this process via affine extensions in subsection \ref{aff}.

\subsection{Architecture}\label{arch}
Though architecture does not necessarily, and does not traditionally, rely on a single kind of identical building block using symmetry (though prefabricated items such as bricks, joints or roof tiles of course feature heavily), the second part of the 20th century saw the rise of the work of architect Buckminster Fuller. His geodesic domes, made of (parts of) spherical tilings in terms of the above triangulations or hexagon-pentagon arrangements (which are dual to each other) found widespread prominence, for instance in the Eden Project, or even the hotel adjacent to the Chern Institute. Similarly, icosahedral design principles found applications in 3D microphones, radar stations and even nuclear warheads. However, as we saw above, the triangulations are not the only (quasiequivalent or not) icosahedral tilings. For instance, the new Amazon headquarters in Seattle feature various such more general tilings, including what is essentially the kite-rhombus tiling from HPV.

\subsection{Root systems and reflection groups}\label{root}
The setting for a {root system} is  an $n$-dimensional Euclidean vector space $V$ endowed with a positive definite bilinear form.
As we will discuss further later, this allows one to construct without loss of generality the (universal) Clifford algebra over this vector space.
The root system is then a collection of non-zero (root)  vectors $\alpha$ spanning $V$ and satisfying the  two axioms. Firstly, that if a root $\alpha$ is contained in it then so is its negative (but no other scalar multiples). 
Each (root) vector defines a reflection in the hyperplane that it is normal to given by the formula $s_\alpha: \lambda\rightarrow s_\alpha(\lambda)=\lambda - 2\frac{(\lambda|\alpha)}{(\alpha|\alpha)}\alpha$. We will assume that the roots are unit vectors. This formula uses the inner product in subtracting twice the component of the vector parallel to the unit normal of the reflection hyperplane, and it will substantially simplify in the Clifford algebra framework. These reflections in the roots can then be multiplied together to form a reflection (or Coxeter) group, such that a root system is a convenient `skeleton' for generating reflection groups. The second axiom is then that the root system be  invariant under all such reflections.
The so-called {simple roots} are sufficient to express every root as a linear combinations with coefficients of the same sign. 
The {crystallographic} root systems that arise in Lie theory via the root lattice have integer linear combinations, whilst for certain {non-crystallographic} root systems one needs to  extend the notion of integer, e.g. allow a  golden ratio component for the groups $H_2\subset H_3 \subset H_4$. 
Usually in Lie theory one might start from a Lie group (e.g. a gauge theory) and then consider the Lie algebra of infinitesimal transformations. The commuting generators (quantum numbers/maximal torus) are in some sense trivial and by going to the Cartan-Weyl basis all the interesting structure of the algebra is found to be given by a root lattice, of which the reflection group is in this context called the Weyl group \cite{Humphreys1990Coxeter, FuchsSchweigert1997}. For the crystallographic root systems one can reverse this process up to integrability conditions of the Lie group/algebra, which however are satisfied by the crystallographic condition. One can affinise the root system into the root lattice by adding affine hyperplanes (again employing the inner product in the definition) that do not contain the origin; combinations of reflections in parallel planes essentially amount to a translation, which forms infinitely copies of the root system in the form of a root lattice.

\subsection{Affine extensions  and lattices}\label{aff}

The  affine root for this procedure is usually (minus) the highest root $\alpha_H$, such that the two reflections are $s_\alpha: \lambda\rightarrow \lambda - 2\frac{(\lambda|\alpha)}{(\alpha|\alpha)}\alpha$ and $s_\alpha^{aff}: \lambda\rightarrow \lambda +\alpha_H - 2\frac{(\lambda|\alpha)}{(\alpha|\alpha)}\alpha$. This gives one well-defined translation by $\alpha_H$ -- the cancellation of the terms is a non-trivial exercise but will become trivial in the Clifford approach -- which gives exactly the right translation to create the root lattice, but other translations might create other lattices. The non-crystallographic root systems by definition have no such lattice. In the usual Kac-Moody procedure the integer entries of the generalised Cartan matrices appear as integer powers of generators in the Chevalley-Serre relations; so it appears that the appearance of the golden ratio in this context is non-sensical. So this translation cannot be allowed to act infinitely many times for the non-crystallographic root systems; but one could still entertain the idea of a meaningful affine extension obtained by finite action of the translation (e.g. once), for instance to describe extended icosahedral structures such as viruses and fullerenes. If one starts with say an orbit of a certain compact group, one could create a translated copy and then take the orbit under the compact group again. The resulting point arrays will have the some overall compact symmetry that one started with, but  the translation creates additional constraints, in particular if the cardinality of the resulting point set is less than maximal. 

Considering such compact point sets derived  via an affine translation given by the highest root was first done in 
\cite{Twarock:2002a} (which amounted to a unit translation along a 2-fold axis).  \cite{Keef:2012, Keef:2009} considered  symmetry systems by letting a translation act  on a starting configuration. \cite{DechantTwarockBoehm2011H3aff, DechantTwarockBoehm2011E8A4} derived such affine extensions in the root system/Coxeter group framework in two ways. Firstly, one can start with the non-crystallographic group, say the icosahedral group $H_3$, and directly try to extend the non-crystallographic group in the above fashion by adding different affine roots. This results in a translation, an extended diagram and Cartan matrix \cite{DechantTwarockBoehm2011H3aff}. This generalised \cite{Twarock:2002a} by considering affine roots/translations of different lengths and in different directions, such as  3-fold and 5-fold symmetry axes. The paper classified the many possible translations into families related by Fibonacci rescaling. Secondly, one can also arrive at affine extensions of the non-crystallographic groups (which might be controversial) by combining two well-known constructions (which are uncontroversial) \cite{DechantTwarockBoehm2011E8A4}. The root systems $A_4\subset D_6 \subset E_8$ are crystallographic and thus have well-defined affine extensions using the highest root. However, there exist projections of the simple roots of  $A_4\subset D_6 \subset E_8$  onto the simple roots of $H_2\subset H_3 \subset H_4$. The crystallographic affine root is a linear combination of the simple roots, e.g. for $E_8$ $$-\alpha_0=2\alpha_1+3\alpha_2+4\alpha_3+5\alpha_4+6\alpha_5+4\alpha_6+2\alpha_7+3\alpha_8,\label{affine}$$ such that the projection of this root can be calculated via linearity of the projection. This then defines an induced affine root for the non-crystallographic root systems. This gave a subset of translations considered in the other construction, along 2- and 5-fold symmetry axes. A related construction of compact non-crystallographic point sets by applying the above projection to orbits in higher dimensions was performed in \cite{twarock2015orbits}.

These constructions give a finite library of blueprints that might be realised by viruses, and from which a best fit can be selected by matching the capsid protein shell; the remainder of the point set is then a prediction for the interior of the virus. Indeed such icosahedral organisation of the RNA has been observed in e.g. MS2 and Pariacotovirus, with an excellent fit \cite{Koning:2003, Janner:2011b, Dechant2012AGACSE, keef2013structural}. In these cases there is an icosahedral component of the RNA distribution that makes contact with the  capsid protein (CP) at locations given by orbits of the icosahedral group.  The idea that symmetry orchestrates also the interior NA distribution of viruses has led to a new principle in how viruses ensure specific encapsidation of their viral RNA (as opposed to abundant cellular RNAs). This will be the subject of the next subsection, whilst application of affine symmetry to fullerenes will be in the following subsection. 

\subsection{A new insight into viral assembly:  Hamiltonian paths on icosahedral solids}\label{ass}
The appreciation of the importance of the RNA-CP contact has led to a step-change in the understanding of viral assembly \cite{Toropova:2008, geraets2015asymmetric, dykeman2014solving, Dykeman:2011, Morton:2010, patel2015revealing, rolfsson2016direct, stockley2016bacteriophage, stockley2013packaging, dykeman2013packaging, stockley2013new}. Essentially the secondary structures of the RNA (stemloops/packaging signals PS) recruit coat proteins during assembly that then get added to the growing capsid. This packaging signal sequence thus helps viral assembly, and  the order of addition of protein building blocks determines a Hamiltonian cycle on the polytope. Moreover,  $60$ of the $90$ rhombuses of the MS2 tiling only turn into the right conformation for the tiling and assembly upon binding the RNA stemloop (allosteric conformational change) \cite{Rolfsson:2010}. The RNA thus has to, in an abstract and idealised way, visit each of the tiles exactly once (so if it is visiting the mid-point of each tile we actually a Hamiltonian path on the dual polyhedron). This is the well-known Hamiltonian path problem on a range of icosahedral solids motivated by the concrete biological systems, i.e.  on or near which symmetry axes the interaction takes place. Additional biological input is that another of the 4 MS2 proteins (maturation protein MP) essentially circularises the RNA, leading to a Hamiltonian (pseudo-)cycle. Furthermore, the cycle gives the order of the addition of capsid proteins and thus the number of bonds formed at each step (at least in this mathematical model), with the creation of more bonds being thermodynamically more favourable. Assembly kinetic considerations therefore allow one to narrow down the cycles ($66$) to the thermodynamically favourable ones. Moreover, clever experiments allow 5-fold averaging or asymmetric reconstructions (normal experimental conditions perform icosahedral averaging, as usually there is no way of identifying individual vertices) of the virion by exploiting the fact that upon infection it binds F-pili of \emph{E. coli} with the distinguished vertex where MP is located (MS stands for male-specific, and is thus a sexually transmitted disease of \emph{E. coli}) \cite{dent2013asymmetric, geraets2015asymmetric}. This allowed to identify a unique Hamiltonian path; moreover, this path appears to be the same in the related GA virus, and thus seems evolutionarily conserved. This strategy thus appears to be highly advantageous to viruses, as it ensures vRNA-specific encapsidation and a much more efficient and  specific assembly process. 

The discovery of this fundamental mechanism via the RNA-CP contact and the PS-mediated assembly instructions that it provides allows further exploration in medicine and nanotechnology. Firstly, one could try to inhibit the RNA or CP binding sites via decoys or other ways of interfering with and misdirecting this assembly process as new anti-viral strategies \cite{patel2017hbv, stewart2016identification}. Secondly, one could fashion virus-like particles based on this design and efficient assembly process, for instance for drug delivery, gene therapy or nanotechnology
\cite{indelicato2016principles, indelicato2017classification}. 

\subsection{Fullerenes}\label{full}
As mentioned above, viruses are not the only naturally occurring icosahedral systems -- a football-shaped configuration of $60$ carbon atoms is well-known as a Buckyball, after Buckminster Fuller. In fact, there is a class of such fullerenes, some icosahedral and some with lesser degrees of symmetry \cite{Marks:1973, Kroto:1985, Kustov:2008}. However, the requirements of carbon chemistry stipulate that typically each carbon atom have three carbon atom neighbours with roughly equal bond lengths and angles. This trivalency requirement rules out the icosahedron as such, but for instance the dodecahedron satisfies this constraint, though the bond angles seem somewhat too extreme to actually occur. Therefore the best-known example is that of the truncated icosahedron $C_{60}$. This configuration has the  pentagons vertex-to-vertex with a hexagon between three pentagons (i.e. on the icosahedral faces), and is thus essentially equivalent to a $T=3$ Caspar-Klug capsid with $h=k=1$. 

One could wonder whether applying our affine extension framework to this system might be useful \cite{Twarock:2002b, Dechant2012AGACSE}; however, we need to take into account the different constraints of carbon chemistry \cite{Dechant:eo5029}. Having generated a point set from making displaced copies of the $C_{60}$, one should therefore look for (outer) shells that satisfy the same requirements that one started with, of trivalency and uniform bond lengths/angles. For the $C_{60}$ this picks out a unique translation along a 5-fold axis that generates a valid shell consisting of $240$ carbon atoms. This has a hexagon inserted vertex-to-vertex between the two vertex-to-vertex pentagons and is thus the counterpart of a $T=12$ Caspar-Klug capsid ($h=k=2$). The next in the series is $C_{540}$ with an additional hexagon and corresponding to $T=27$. This translation is thus a general procedure to derive larger and larger cages indexed by integer $h=k$ and $T=3h^2$ from the starting $C_{60}$. In fact, these cages can exist inside each other in a nested `Russian doll' configuration, also called carbon onion. Various carbon onions have been observed in nature \cite{Ugarte:1992, Ugarte:1995, Kroto:1992}, and it is remarkable that the whole configuration overall can be described by the affine extension procedure. Another icosahedral configuration is $C_{80}$, which has a hexagon edge-on between two pentagons (i.e. along the icosahedral edges), corresponding to $T=4$ with $h=2, k=0$. Again a translation along the 5-fold axis is picked out uniquely that inserts hexagons at each step, generating a carbon onion $C_{80}-C_{180}-C_{320}\dots$, which is indexed by integer $h$, vanishing $k$, and the square numbers $T=h^2$. It is  remarkable that affine symmetry, perhaps with the translation length determined by some inherent length scale in the  problem, such as the bond length of carbon, might occur in nature, both in viruses and fullerenes. 

\subsection{Vibrations of icosahedral objects and representation theory}\label{vib}

Group theory has long been used to calculate vibrational frequencies of symmetric arrangements in physics and chemistry. By Taylor expanding the potential to quadratic order around an equilibrium point (hence the linear term vanishes), one can find simple harmonic oscillator solutions for small vibrations around this equilibrium; if the configuration is symmetric, these normal modes are given by the representation theory of the symmetry group. One first constructs the characters of the permutation representation, which count how many vertices of the configuration are invariant under the various conjugacy classes. One then takes the product with the characters of a 3-dimensional irreducible representation to construct the characters of the displacement representation. This character system can then be decomposed into sums of the character systems of the irreducible representations, thus decomposing the displacement representations in terms of irreducible representations. 

It is not surprising that such a standard technique is also applicable to icosahedral systems such as viruses and fullerenes.
\cite{Peeters:2009, Englert2008twenty} performed such a top-down coarse-grained group theoretical analysis, whilst all-atom simulations were performed in \cite{Dykeman:2010}. We here give the decomposition of various displacement representations of icosahedral solids. E.g. the icosahedron has $12$ vertices on the $5$-fold axes. Thus $5$-fold rotations each fix $2$ points whilst no other conjugacy class fixes any, except for the identity, which of course leaves all $12$ invariant. The decomposition then yields  $\Gamma^{1} + 3 \Gamma^{3} + 1 \Gamma^{3'} +2\Gamma^{4}+3\Gamma^{5}$, which gives the correct total dimension of $36=12\cdot 3$. The dodecahedron has its $20$ vertices on the $3$-fold symmetry axes, such that those 3-fold rotations leave two points invariant; however, the corresponding character in the $3$ irrep is $0$, such that the character system is zero except for the identity, which is $20\cdot 3=60$. Thus the decomposition is of course $\Gamma^{1} + 3 \Gamma^{3} + 3 \Gamma^{3'} +4\Gamma^{4}+5\Gamma^{5}$, analogous to the computation of the allowed dimensions of the irreps in the first place and their sum being equal to the order of the group $1^2+3^2+3^2+4^2+5^2=60$. Something analogous happens to icosahedral solids that have no points on symmetry axes, such as the truncated icosahedron, the truncated dodecahedron or any Caspar-Klug capsid. The corresponding representation will therefore just be an appropriate multiple of the above, e.g. $T(\Gamma^{1} + 3 \Gamma^{3} + 3 \Gamma^{3'} +4\Gamma^{4}+5\Gamma^{5})$ giving total dimension $T\left( 1\cdot 1 + 3\cdot 3 +3\cdot 3+4\cdot 4 + 5 \cdot 5\right)= 60T$. Thus, the following cases with points on symmetry axes remain interesting. The icosahedral root system $H_3$, the icosidodecahedron, has $30$ vertices on the $2$-fold axes, and its decomposition is $\Gamma^{1} + 5 \Gamma^{3} + 5 \Gamma^{3'} +6\Gamma^{4}+7\Gamma^{5}$. Its dual, the above-mentioned rhombic triacontahedron has vertices on the 3- and 5-fold axes and  has decomposition $2\Gamma^{1} + 6 \Gamma^{3} + 4 \Gamma^{3'} +6\Gamma^{4}+8\Gamma^{5}$. The MS2 tiling has $4$ vertices in the fundamental domain, including the 3- and 5-fold axes, making a total of $152$ vertices. The decomposition is thus $8\Gamma^{1} + 24 \Gamma^{3} + 22 \Gamma^{3'} +30\Gamma^{4}+38\Gamma^{5}$. Again, there is some ambiguity whether the relevant quantity is not the position of the vertex but of the centre of mass of the tile, and thus whether the decomposition of the displacement representation of the dual polytope is the relevant one. For instance, \cite{Peeters:2009, Englert2008twenty} have shown that generically Caspar-Klug viruses have $24$ low frequency modes.

\section{Clifford algebras, exceptional root systems and ADE correspondences}\label{partii}
Having worked with (non-crystallographic) root systems, reflection groups and ADE-type (affine) Lie algebras up to now, we now argue why Clifford algebras are a very natural, and powerful, framework in this context. Firstly, for the usual reflection formula one subtracts twice the component parallel to the hyperplane normal, and normally (e.g. in the standard Weyl reflection form) this is given by the inner product. The definition of a root system explicitly stipulates that the setting be a vector space with an inner product. The (without loss of generality, universal) Clifford algebra over that vector space using this bilinear form is therefore an extremely natural and general object to consider in this context (if not \emph{the most natural}) \cite{Dechant2015ICCA}. Secondly, the reflection formula that Clifford algebra allows is extremely simple and powerful, in particular as most transformations of interest can be written as products of reflections (Cartan-Dieudonn\'e theorem). In fact, the reflections and thus also the other transformations are doubly covered in this setup, which constitutes a very simple construction of the Spin groups, and makes an immediate connection with the second root system axiom. Thirdly, even over a real vector space setting, the Clifford algebra contains many geometric objects such as planes that have complex or quaternionic algebraic properties on top of their geometric meaning, which makes complexification unnecessary and instead offers deeper insight into the geometry. After introducing these basics, we will briefly discuss some of the new results found with this approach, such as the geometry of the Coxeter plane, induction constructions of root systems, for $E_8$ and the exceptional 4D root systems, respectively (subsections \ref{E8} and \ref{spinor}), and various (new and old) ADE correspondence that they all culminate in (subsection \ref{ADE}).

\subsection{Clifford algebras}\label{clifford}

Clifford algebras \cite{Hestenes1966STA,  LasenbyDoran2003GeometricAlgebra, Porteous1995Clifford, Lounesto1997, Garling2011Clifford} 
have been used for a very long time as a framework that offers deep geometric insight into mathematical problems. And since physics is essentially geometry, it has shed light on many physical problems from electromagnetism, to special relativity and quantum mechanics. In the root system and reflection group context, however, the elegant reflection formula comes in very useful \cite{Hestenes1990NewFound,Hestenes2002PointGroups,Hestenes2002CrystGroups,Hitzer2010CLUCalc}, and the double cover property connects with the root system axioms \cite{Dechant2012Induction}. So by combining both frameworks, we have been able to get new insights into the geometry as well as new results.

One uses the vector space with inner product structure to construct the Clifford algebra by defining a product  for the vectors  via  $xy=x\cdot y+x \wedge y$. The inner product (given by the symmetric bilinear form) is the symmetric part $x\cdot y=(x|y)=\frac{1}{2}(xy+yx)$ and the exterior product the antisymmetric part $x\wedge y=\frac{1}{2}(xy-yx)$, such that parallel vectors commute whilst orthogonal vectors anticommute. This extends the algebra  to a $2^n$-dimensional vector space (akin to the exterior algebra), and we extend the definition of the product to the whole algebra via linearity and associativity (unlike the exterior algebra). Using this form for the inner product in the reflection formula one gets the simplified Clifford version 
$$s_i: x\rightarrow s_i(x)=-\alpha_ix\alpha_i,$$ since we can now multiply vectors (in particular root vectors) together in this algebra. 
In fact by the Cartan-Dieudonn\'e theorem one can write a large class of transformations as products of reflections \cite{Dirac1936,HestenesSobczyk1984,Dechant2015ICCA}.
Thus Clifford algebra actually affords a completely general way of performing such transformations by successive multiplication with unit vectors defining the reflection hyperplanes
$$s_1s_2\dots s_k: x\rightarrow s_1s_2\dots s_k(x)=(-1)^k\alpha_1\alpha_2\dots\alpha_kx\alpha_k\dots\alpha_2\alpha_1=:\pm Ax\tilde{A}.$$
(The tilde denotes the reversal of the order of the constituent vectors in the product.) The products of root vectors in the Clifford algebra therefore give group transformations, so Clifford algebra provides a new way for doing group theory for such groups.

Since $n$ and $-n$ doubly cover the same reflection, products of unit vectors doubly cover the respective orthogonal transformation. We call even products, i.e. products of an even number of vectors, `spinors' $R$, and a general product $A$ `pinors'. The latter pinors form the $\Pin$ group and constitute a double cover of the orthogonal group, whilst the even products (spinors) form the double cover of the special orthogonal group, called the $\Spin$ group.

The above features are completely general for orthogonal groups in spaces of arbitrary dimension and signature. In particular, the `single-term' quaternion reflection formula is often seen as very deep -- but it is only accidental that $\Spin(3)$ is isomorphic to the quaternions, whilst the Clifford formula is completely general.  As an illustrative example (and because we will be concerned with the accidentalness of 3D and its connections with 4D and 8D) we consider the Clifford algebra of 3D generated by three orthogonal unit vectors $e_1$, $e_2$ and $e_3$. This yields an eight-dimensional vector space consisting of one scalar $1$, three vectors $e_1, e_2, e_3$, three bivectors $e_1e_2, e_2e_3, e_3e_1$ which square to $-1$ and describe planes, and one trivector $e_1e_2e_3$. 
 Thus, one gets different imaginary units based on a real vector space, without complexifying the whole space.  In fact, the scalar and the three bivectors also satisfy quaternionic relations. Thus, quaternions arise in many contexts where geometrically the group $\Spin(3)$ is at work, such as quaternionic representations of root systems or representations of quaternionic type of the polyhedral groups. This is an underlying property of space, rather than a special feature of each individual group, which seems to have been consistently overlooked in the literature. Thus, one sees that the geometry of three dimensions is intimately related to the geometry of  four and eight dimensions.

\subsection{$E_8$ from the icosahedron}\label{E8}
$E_8$ is a root system in eight dimensions, consisting of $240$ vertices. We have seen earlier that $H_3$ is a 3D reflection group with $120$ elements. Now using the Clifford way of multiplying together 3D root vectors to give a double cover of this group, we get $240$ elements of the 8D Clifford algebra. So far so intuitive that this will give $E_8$. However,  with a generic inner product this just gives two copies of $H_4$. Conway knew (though first  published in \cite{Wilson1986E8}) how to turn two copies of $H_4$ into the $E_8$ root system by using a reduced inner product. However, it does not seem to have been known that one can get $H_4$ from $H_3$ when considering the double cover of $A_5$, and even two copies of $H_4$ when considering the double cover of $H_3$ (i.e. including the inversion), which allows one to perform the $E_8$ construction \cite{Dechant2016Birth}. 

\subsection{3D to 4D spinor induction}\label{spinor}

This construction of $E_8$ from $H_3$ is beautiful and deep on the one hand; on the other it is a single case and the use of the reduced inner product is somewhat ugly. There is a very beautiful and elegant general statement about the relationship between root systems in 3D and 4D, that we will discuss in this section \cite{Dechant2012Induction, Dechant2012CoxGA}:
 any 3D root system yields a 4D root system, and thereby a corresponding reflection group. 

Multiplying root vectors in the Clifford algebra generally yields pinors in the full 8D algebra, but even products will stay within the even 4D subalgebra of scalar and bivectors. They will be of the general form $R=a_0+a_1e_2e_3+a_2e_3e_1+a_3e_1e_2$. Endowed with the inner product  $(R_1,R_2)=\frac{1}{2}(R_1\tilde{R}_2+R_2\tilde{R}_1)$, this is actually just four-dimensional Euclidean space with the norm $|R|^2=R\tilde{R}=a_0^2+a_1^2+a_2^2+a_3^2$.

 The polyhedral groups act on a vector $x$ via $\tilde{R}xR$ with the spinors $R$ forming a spin double cover; in fact these spinors constitute the respective binary polyhedral groups with  multiplication $R_1R_2$.   Thus a group of spinors directly  yields a collection of vectors in 4D, which are easily shown to be root systems: the first axiom is trivially satisfied by the double cover property (this holds in fact in any dimension); the closure under reflections with respect to the above inner product is also easily shown due to properties of the spinor groups, but do not (easily) generalise to other dimensions. The statement is thus that
any 3D root system gives rise to a spinor group $G$ which induces a root system in 4D.
 The accidentalness of this construction helps explain the abundance of root systems in four dimensions due to  additional, exceptional root systems. The set of irreducible 3D root systems
$(A_3, B_3, H_3)$ gives rise to the exceptional set $(D_4, F_4, H_4)$ in 4D. For the reducible ones $A_1
\times I_2(n)$ one gets a doubling $I_2(n)\times I_2(n)$. There are analogues in 2D and with the octonions; however, the latter just give two copies of the above case whilst the 2D family is simply self-dual: in 2D the set of vectors $\{e_1, e_2\}$ is just Hodge dual to the set of spinors $\{1, e_1e_2\}$, so this case does not look terribly interesting at this stage. 

\subsection{Trinities and the Coxeter plane}\label{Trinities}
The sets $(A_3, B_3, H_3)$ and $(D_4, F_4, H_4)$ are examples of a recurring pattern in Mathematics: three exceptional cases (Trinities) -- and sometimes two infinite families (ADE-type sets). The great mathematician Vladimir Arnold noticed about $30$ Trinities throughout mathematics, sometimes with clear connections, sometimes esoteric ones, sometimes none at all except via the similarity or perhaps multiple intermediate steps \cite{Arnold1999symplectization,Arnold2000AMS}.

Arnold found a rather indirect link between $(A_3, B_3, H_3)$ and $(D_4, F_4, H_4)$, about which he says ``\emph{Few years ago I had discovered an operation transforming the last trinity [$(A_3, B_3, H_3)$] into another trinity of Coxeter groups $(D_4, F_4, H_4)$. I shall describe this rather unexpected operation later.}'' His Weyl chamber Springer cone decomposition of $(A_3, B_3, H_3)$ decomposes the  orders of the groups as
$24=2(1+3+3+5)$, 
$48=2(1+5+7+11)$ and 
$120=2(1+11+19+29)$, which he notices are coefficients that are one less than the quasihomogeneous weights of $(D_4, F_4, H_4)$, which are $(2, 4, 4, 6)$, $(2, 6, 8, 12)$ and  $(2, 12, 20, 30)$, respectively. 

These numbers are actually more directly the exponents $m_i$ of $(D_4, F_4, H_4)$, which are well known to be one less than the degrees of polynomial invariants \cite{Humphreys1990Coxeter}.  That is, they are given by the complex eigenvalues $\exp(2\pi i m_i/h)$ of the Coxeter element and $d_i=m_i+1$ for degrees $d_i$. This streamlines Arnold's observation somewhat, but is still not much more than a suggestive link, which my  Clifford construction establishes explicitly for the first time. In fact, this construction also contained other cases given by the infinite family  $A_1\times I_2(n)$ giving $I_2(n)\times I_2(n)$. One can thus wonder whether Arnold's observation in terms of the decomposition of the groups and the exponents might extend to include these additional cases. Indeed, it turns out that they do, which further establishes the connection between $(A_1\times I_2(n), A_3, B_3, H_3)$ and $(I_2(n)\times I_2(n), D_4, F_4, H_4)$. 

This calculation for the 2D cases is somewhat trivial, but involves finding the exponents of the Coxeter element. The usual procedure is to complexify the vector space and looks for complex eigenvalues. Complex eigenvalues of course mean the vector is not an eigenvector at all, but gets rotated in some plane and only returns back to itself after several applications. So it is more natural to think of the plane as an eigenplane rather than of eigenvectors. In fact, in Clifford algebra rotations in planes are described by bivector exponentials, e.g. reflections in the vectors $\alpha_1=e_1$ and $\alpha_2=-\cos{\frac{\pi}{n}}e_1+\sin{\frac{\pi}{n}}e_2$ actually amount to an $n$-fold rotation given by 
\begin{equation}
W=\alpha_1\alpha_2=-\cos{\frac{\pi}{n}}+\sin{\frac{\pi}{n}}e_1e_2=-\exp{\left(-{\pi e_1e_2/n}\right)}.
\end{equation}

Thus, no complexification is necessary, as the bivectors square to $-1$ but give the rotation plane algebraically  \cite{Dechant2017e8,Dechant2012AGACSE}. So say that we can completely factorise the pinor $W=\alpha_1 \dots \alpha_n$ which gives the action of the Coxeter element in terms of commuting factors $\tilde{W}vW=\tilde{W_1}\dots\tilde{W_k}vW_k\dots W_1$ with ${W_i}=\exp(\pi B_im/h)=\cos (\pi m/h) + \sin (\pi m/h) B_i$ that are bivector exponentials of a bivector $B_i$ describing a certain plane. Let us consider a vector $v$ that is orthogonal to $B_i$. Then it is also orthogonal to two vectors that generate $B_i$ upon multiplication; so it anticommutes with both, such that it commutes with $B_i$. So actually $W_i$ commutes with $v$ such that $v\rightarrow \tilde{W_i}vW_i=\tilde{W_i}W_iv=v$ and the vector is invariant under rotation in that plane. However, if $v$ lies in the plane defined by $B_i$, it will be orthogonal to one of the vector factors but parallel to the other; thus, $v$ now anticommutes with $B_i$. 
So when taking $W_i$ through to the left now one introduces a minus sign in  the bivector part (but not the scalar part). This is equivalent to reversal $\tilde{W_i}$ and one therefore gets the usual complex eigenvector equation 
$
	v\rightarrow \tilde{W_i}vW_i=\tilde{W_i}^2v=\exp{\left(\pm{2\pi m_i B_i/h}\right)}v
$%
 {} without the need for complexification, because the complex structure simply arises from the bivector describing the rotation plane.  
So for a Coxeter pinor $W$ factorisation into orthogonal eigenspaces $W=W_1\dots W_k$  as above,  all the orthogonal $W_i$s commute through and cancel out, whilst the one that defines the eigenplane that $v$ lies in  gives the  complex eigenvalue equation with respect to this plane.
Now all the exponents arise purely algebraically from the factorisation of the product of simple roots in the Clifford algebra  just as righthanded and lefthanded rotations in the respective eigenplanes. E.g. $\alpha_1\alpha_2\alpha_3\alpha_4$ for $H_4$ factorises as $\exp\left(-\frac{\pi}{{30}}B_C\right)\exp\left(-\frac{{11}\pi}{30}e_1e_2e_3e_4B_C\right)$ for the Coxeter plane $B_C$, giving exponents $1,11,19,29$ directly from the simple roots.

\subsection{McKay correspondence and ADE}\label{ADE}
We have shown above that there is a direct link between $H_3$ and $E_8$, as well as between the 3D and 4D root systems via the intermediate spinor groups. They are even subgroups of $SU(2)$, by virtue of $\Spin(3)$ being isomorphic to $SU(2)$. In order to go beyond the above Trinities of three exceptional cases to include the infinite families let us first review the McKay correspondence \cite{Mckay1980graphs}, named after John McKay who also first noticed Moonshine \cite{gannon2006moonshine,eguchi2011notes}. It relates the binary polyhedral groups $(2T, 2O, 2I)$ and the Trinity of $E$-type Lie algebras $(E_6, E_7, E_8)$ in the following two ways: firstly, the graph that can be constructed from the tensor product structure of the irreducible representations of each binary polyhedral group is exactly the  affine  $(E_6, E_7, E_8)$ graph (c.f. also Section \ref{viruses} for the binary icosahedral group as well as Eq. (\ref{affine}) for the affine root of $E_8$). Secondly,  the sum of the dimensions of the irreducible representations  $(12, 18, 30)$ is equal to the Coxeter number $h$  of the $E$-type Lie algebras $(E_6, E_7, E_8)$. In fact the McKay correspondence is wider than just the Trinity, and contains all subgroups of $\Spin(3)$, including the binary cyclic and dicyclic groups along with the binary tetrahedral, octahedral and icoshedral groups, and the $ADE$ Lie algebras. 

So we can connect the two constructions: use spinor induction  from $(A_3, B_3, H_3)$ and then the McKay correspondence to arrive at $(E_6, E_7, E_8)$ via the $SU(2)$ intermediates. Of course we had already shown the direct connection for $H_3$ and $E_8$. The first puzzling thing to note is that the McKay connection $(12, 18, 30)$ in terms of irreducible representations and Coxeter numbers is also just the number of roots in $(A_3, B_3, H_3)$. Secondly, our construction contained $A_1\times I_2(n)$, which in the same way can be matched up with the $D_n$ family. Further, if we want to connect with the $A_n$ cases we should now also include the self-dual 2D root systems $I_2(n)$ in the correspondence, which appeared somewhat trivial at first. Again the Coxeter numbers and numbers of irreps in the McKay case match up with the number of roots in $(I_2(n), A_1\times I_2(n), A_3, B_3, H_3)$, which is very puzzling and hints that the correspondences are all rooted in these polyhedral root systems. There is thus a three-way chain of correspondences between 
$(I_2(n), A_1\times I_2(n), A_3, B_3, H_3)$, $(I_2(n), I_2(n)\times I_2(n), D_4, F_4, H_4)$ and $ (A_n, D_n, E_6, E_7, E_8)$. These are all root systems, although we tend to think of the first as polytopes, the second as subgroups of $SU(2)$ and the third as Lie algebras or groups -- however, to our knowledge, all the structure is already contained at the level of the root system and reflection group, and does not require the Lie theoretic context. There thus appears to be a conceptual unification at the level of root systems for all three sets and the correspondences among them.
 
It is therefore intriguing whether there is a direct correspondence between $(I_2(n), A_1\times I_2(n), A_3, B_3, H_3)$ and $ADE$ without involving the intermediate $SU(2)$, e.g. generalising the $H_3$ to $E_8$ construction. Furthermore, not all Lie algebras of $A$-type are actually covered in this setup because root systems are always even, whilst the odd order cyclic groups also correspond to $A$-type Lie algebras, which are therefore not covered (one could argue that one could also consider the cyclic groups of order $n$,  $C_n$, since they are subgroups and implicitly contained). However, that would lose our root system based reasoning which proved so fruitful for the spinor induction constructions. We therefore seek a direct correspondence between $(I_2(n), A_1\times I_2(n), A_3, B_3, H_3)$ and $ADE$. The recent paper by He and McKay \cite{he2015sporadic} presented a connection between $(A_3, B_3, H_3)$ and  $(E_6, E_7, E_8)$ of the perhaps more esoteric-appearing kind: the triples of rotational orders generated by the polyhedral groups is identical to the triple of the lengths of the legs of the $E$-type diagrams. E.g. the triple for the icosahedral group is $235$ since there is $2$-fold symmetry around the edges, $3$-fold symmetry about each face and $5$-fold symmetry around each vertex (alternatively these are the angles between simple roots as fractions of $\pi$). Equivalently these are the (suppressed) labels in the Coxeter-Dynkin diagrams: two orthogonal roots have $2$, a simple link is 3, and the final link is labelled 5. For $E_8$ the legs, including the central node in each, have lengths $2$, $3$ and $5$. This holds for the Trinities: $(233, 234, 235)$ describes $(A_3, B_3, H_3)$ and  $(E_6, E_7, E_8)$. Now, motivated by the above inclusion of the two infinite families we hypothesise a full correspondence for $(I_2(n), A_1\times I_2(n), A_3, B_3, H_3)$: indeed, one has that $I_2(n)$ just generates one $n$-fold rotation which gives a single leg of length $n$, which is thus the $A_n$ diagram; $A_1\times I_2(n)$ has two order $2$ rotations from the $A_1$ node with either of the $I_2(n)$ nodes, whilst the latter together form an $n$-fold rotation. This gives a diagram with leg lengths $2$, $2$ and $n$ -- which is just the diagram for $D_{n+2}$. Thus, based on the above considerations for what the correspondence should be one gets exactly all the right $ADE$ diagrams. This direct connection is puzzling, and it would be nice to make this connection more explicit in due course. Thus there are actually three-way $ADE$ correspondences between the three different sets of root systems $(I_2(n), A_1\times I_2(n), A_3, B_3, H_3)$, $(I_2(n), I_2(n)\times I_2(n), D_4, F_4, H_4)$ and $(A_n, D_n, E_6, E_7, E_8)$. It would be nice to show whether this Trinity of sets of root systems and $ADE$ correspondences is determined by a mysterious underlying concept, or whether the two higher-dimensional sets are simply determined by the Platonic root systems. 

\section{Conclusions}\label{concl}

In this paper, we discussed historical and recent applications of group theory to virology. Viral symmetries motivated the study of non-crystallographic groups in the guise of the icosahedral group, and generalised non-compact symmetries motivated the root system and affine extension framework. The ordered NA component inside the capsid and the crucial role of the NA-CP contact led to a paradigm shift in virus assembly with potential applications in antiviral therapies, virus-like particles and nanoparticles for drug delivery. Another application of affine symmetry to fullerenes describes entire carbon onions with different shells being related via the translation operator. 

In the second half of the paper, we argued the case for a Clifford algebra framework, as root systems and reflection groups already provide all the required structure of a vector space with an inner product. This is a very natural and powerful approach, in particular due to the simple reflection formula that yields double covers of group transformations and deeper insight into the geometry, e.g. that of the Coxeter plane. We discussed root system induction constructions from $H_3$ to $E_8$ and $(A_1\times I_2(n), A_3, B_3, H_3)$ to $(I_2(n)\times I_2(n), D_4, F_4, H_4)$, widening Arnold's Trinity and making a constructive link for the first time. The McKay correspondence further motivates the inclusion of $I_2(n)$ in the correspondence, leading to a Trinity of correspondences between the three sets of root systems $(I_2(n), A_1\times I_2(n), A_3, B_3, H_3)$, $(I_2(n), I_2(n)\times I_2(n), D_4, F_4, H_4)$ and $(A_n, D_n, E_6, E_7, E_8)$, e.g. with irreducible representations and Coxeter numbers being determined by the number of roots in $(I_2(n), A_1\times I_2(n), A_3, B_3, H_3)$,  whilst the $ADE$-type diagrams are derivable from orders of the generated rotations.

\subsection*{Acknowledgements}\label{ack}
I would  like to thank the organisers of the Nankai Symposium on Geometry, Physics and Number Theory for an excellent meeting and the Chern Institute at Nankai and the Yau Institute at Tsinghua for their hospitality. In particular, I would also like to thank Yang-Hui He, Terry Gannon, John McKay, Peter Cameron, Alastair King, Rob Wilson, Jim Humphreys, Reidun Twarock, C\'eline Boehm, Anne Taormina, David Hestenes, Anthony Lasenby and Eckhard Hitzer for helpful discussions. 
% ------------------------------------------------------------------------

\bibliography{virus,virobib,Moonshine}

% ------------------------------------------------------------------------
\end{document}